\documentclass[twocolumn,secnumarabic,amssymb, nobibnotes, aps, prd]{revtex4-1}

\newcommand{\lr}[1]{\left({#1}\right)}
\usepackage[hidelinks]{hyperref}
\usepackage{float}
\usepackage{svg}

\begin{document}

\title{Chaotic Dynamics of a Dripping Water Faucet}%

\author{Michael Sekatchev}%
\email[Email: ]{michaelsekatchev@live.ca}
\affiliation{Department of Physics and Astronomy, UBC, Vancouver, BC}
\date{November 27, 2022}%

\begin{abstract}
An experimental approach is taken to study the dynamics of the dripping water faucet, a simple deterministic system. The time interval between successive drops may be affected by the many drops preceding it. The time interval is predicted by numerical simulations to exhibit increasingly chaotic behavior with increasing flow rate, showing transitions from single period to period-m ($m=2,4,8,...$) dripping, followed by a purely chaotic regime. Deterministic regimes are identified through plots of the time interval against drop number and against successive time interval, and through a bifurcation diagram, but the dripping faucet does not traverse them sequentially as anticipated numerically. Understanding the chaotic dynamics of the dripping faucet will aid in the study of more complex chaotic systems, of which there are a plethora across many disciplines.
\end{abstract}

\maketitle


The study of chaotic systems was developed by Henri Poincaré, in his investigation of orbits of more than three bodies in the solar system \cite{HOLMES1990137}. Poincaré showed that some dynamical systems are highly sensitive to initial conditions, such as the double pendulum \cite{1993AmJPh..61.1038L}, the orbits of three-body planetary systems \cite{Rath_2022}, or a dripping water faucet \cite{chaos_manual_paper}. The dynamics of these systems are not entirely random as was previously thought; they are characterized by certain underlying complex yet deterministic patterns, including feedback loops, fractals, self-similarities. 

This letter investigates one of the simplest chaotic systems---the dripping water faucet,  first studied in \cite{shaw_original}. The rate of water droplet formation does not remain periodic at all flow rates; The system may follow a period-2 regime, in which successive time intervals between drops $T_n$ alternate between two periods, (with sequence $T_1, T_2, T_1, T_2, ...$), which doubles into a period-4 regime at higher flow rate. At even higher flow rates the system transitions from these complex period-$m$ dripping regimes ($m=2,4,8,...$) to purely chaotic responses in which the successive drop time intervals exhibit complex deterministic relations. The transitions between these dynamical regimes can be studied in a bifurcation diagram, which plots the evolution of these successive intervals $T_n$ against flow rate. 

Since the first pioneering work on the leaky faucet by \cite{shaw_original}, many studies involving experiment and computation have been performed to understand the dynamics and transitions of the system \cite{chaos_manual_paper,prev_chaos_work_cambridge, route_to_chaos,dinnocenzo_1996_dripping,drip_jet_lit_review,japanese_paper}. The Feigenbaum constants, associated with the ratio of the transitions between successive period-$n$ regimes, have been calculated from bifurcation diagrams created from models of a dipping water faucet. 

A detailed study of the dripping water faucet is presented here, which documents the results from an attempt to obtain an estimate of one of the Feigenbaum constants from an experimental setup. This is accomplished by recording the periods of successive water droplets, and sampling the data to produce a bifurcation diagram of the droplet period against the flow rate, scaled to adjust for varying droplet size. Plots of the relationship between successive droplets ($T_n$ v.s. $T_{n+1}$) are also presented for different regimes to show the deterministic behavior of the system.

Chaotic systems exist in many natural systems (fluid dynamics, weather patterns, heartbeat irregularities, etc.) and have applications across a multitude of disciplines (meteorology, economics, sociology, etc.). Understanding the exact chaotic dynamics of a dripping water faucet, one of the simplest chaotic systems with a single variable (the droplet period), can provide valuable insight into the study of these more complex systems.

Feigenbaum discovered two mathematical constants, $\delta$ and $\alpha$, describing the proportions in a bifurcation diagram for any non-linear map \cite{feigenbaum}. In other words, he made the discovery that all bifurcation diagrams, and thus all systems exhibiting period-doubling bifurcation, are analogous to one another. Within the context of the dripping faucet experiment, $\delta$ is the limit of the ratio of the width of one period doubling regime to the next. Mathematically, this is expressed as
\begin{equation}
    \delta = \lim_{m\to\infty}\lr{\frac{F_{m-1}-F_{m-2}}{F_m-F_{m-1}}}=4.669 ...,
\end{equation}
where $F_m$ represents the value of the volumetric flow rate at which the period doubles for the $m$-th time. It is important to make the distinction between volumetric flow rate (ml per min) and droplet flow rate (drops per min), since the shape and volume of droplets has a non-linear dependence on the flow rate \cite{drip_jet_lit_review}. The idea is to work directly with the control parameter, which is the volume of water flowing through the faucet, rather than the drop count. Using the droplet flow rate rather than the volumetric flow rate would therefore not provide the correct spacing in a bifurcation diagram, and make the results incomparable with existing numerical simulations which are created as a function of the control parameter, such as in \cite{dinnocenzo_1996_dripping}. 

\begin{figure}[H]
    \centering
	\includegraphics[width=0.75\columnwidth]{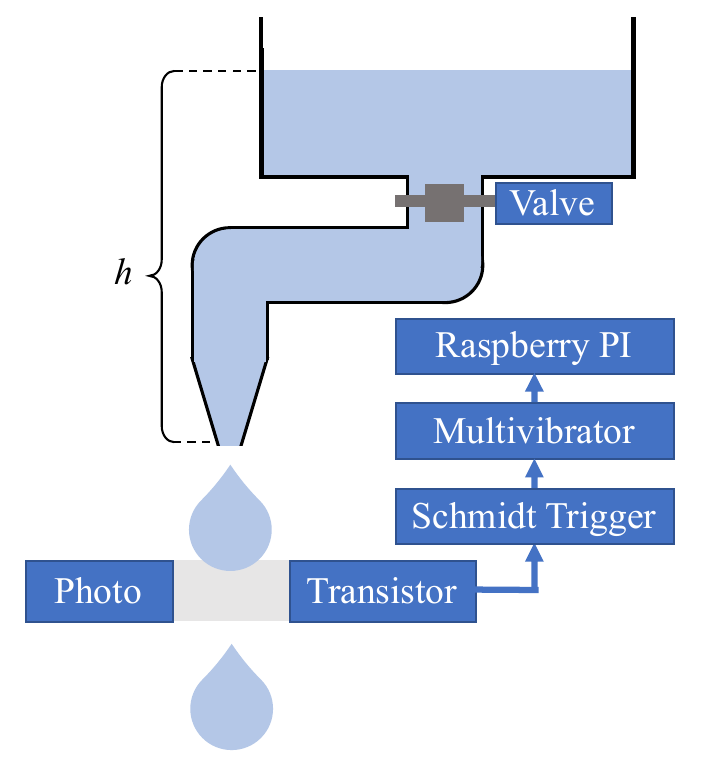}
    \caption{Diagram of the experimental setup used to study the dynamic regimes of a dripping water faucet. A $50\times40\times60$ cm container is filled with water and allowed to drain by dripping through an eyedropper. A phototransistor is positioned below the eyedropper, which emits an electric pulse when a droplet passes through it. This signal actives a Schmidt trigger, which removes noise from the analog phototransistor input. The output from the Schmidt trigger is fed to a one-shot multivibrator, which converts the signal into a simple digital pulse that is read by a connected Raspberry PI. The resulting output is the timestamp $t_n$ of each droplet. The height $h$ between the water level and the eyedropper is the main physical parameter in this experiment, and is directly proportional to the flow rate. A valve is also included for additional flow rate control.}
    \label{fig:setup_diag}
\end{figure}

The second constant, $\alpha$, is the limit of the ratios between the separation of two branches in a period doubled regime and the next,
\begin{equation}
    \alpha = \lim_{m\to\infty}\lr{\frac{\Delta T_m}{\Delta T_{m+1}}}= 2.502...,
\end{equation}
where $\Delta T_m$ refers to the distance between two branches in a period-$m$ regime. Contrary to the first constant $\delta$, Because the second constant is affected by the vertical distances in a bifurcation diagram (of $T_n$ against flow rate $F$), using a droplet flow rate would still allow $\alpha$ to be determined, since the scaling from a droplet flow rate to a volumetric flow rate only affects the horizontal spacing. An experimental setup would involve a precise measurement of the time $t_n$ of each drop, from which the period between successive drops and the flow rate could be calculated. A diagram and description of the experimental setup is provided in Fig. \ref{fig:setup_diag}.


From the raw drop timestep readings $t_n$ obtained from the phototransistor, the time interval $T_n$ between successive drops can be calculated as $T_n = t_n - t_{n-1}$. For adjusting the flow rate to probe different dynamics of the system, a stepper motor with a valve could be used to control the flow rate from the tank. However, simply allowing the tank to drain without replenishing the water supply allows the system to traverse the entire range of available dynamics of a dripping water faucet at a continuously decreasing flow rate (increasing period). The complication with this method is that it introduces a systematic drift to the data, which needs to be accounted for carefully during the analysis, as will soon be discussed. 

\begin{figure}[!]
	\includegraphics[width=\columnwidth]{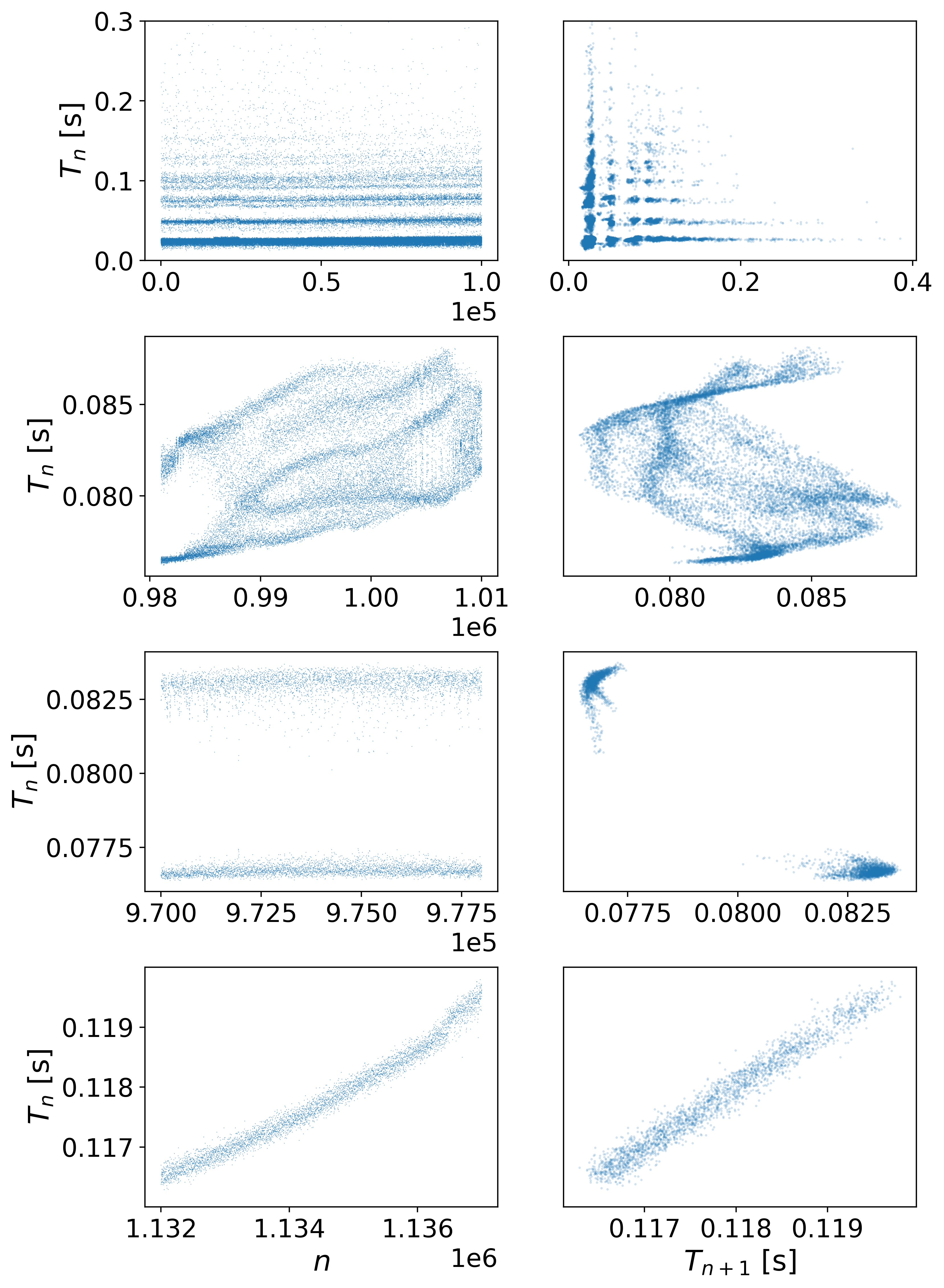}
    \caption{Left column: time interval of successive water droplets from a dripping faucet, plotted against the drop number $n$. Right column: relationship between each successive drop interval $T_n$ and the interval succeeding it, $T_{n+1}$. A container filled with water was allowed to drain by dripping through an eyedropper. The time of each drop was recorded following their detection using a phototransistor. While draining, the tank probed the range of dynamics of a dripping water faucet at a continuously decreasing flow rate (increasing period). Shown are four selected ranges of data which demonstrate the system evolving through four different types of dynamics. Looking from top to bottom, we first observe a missing droplet effect due to the limitations of the phototransistor, which was estimated to have a limit of approximately 0.025 s. For intervals $T_n\lessapprox0.025\,\text{s}$, as is the case here, occasionally every second (or third, or fourth, etc.) drop was recorded, resulting in the vertical bands shown in the $T_n$ v.s. $n$ plot and gaps in the $T_{n}$ v.s. $T_{n+1}$ plot. 
    The second row demonstrates period-4 dripping. Four branches are clearly visible following the transition. The associated ``cobra''-like shape on the $T_n$ v.s. $T_{n+1}$ plot is an indication that there is a deterministic relationship between more than simply two adjacent droplet intervals.
    The third row shows a period-2 regime. The two distinct successive time intervals are seen on the $T_n$ v.s. $T_{n+1}$ plot.
    The last row shows a simple periodic dripping. The successive intervals $T_n$ and $T_{n+1}$ are directly related, changing only linearly with time as the tank empties---A systematic linear drift is observed throughout these plots. The system is observed to change through these four dynamics multiple times within its entire observed evolution.}
    \label{fig:tn_vs_tn1}
\end{figure}

It is exactly in this way, by letting the tank to drain itself, that data was collected. The tank was allowed to drain from full ($50$ cm) to empty in $\sim$2.5 days, during which $1.17\times10^6$ water droplets were recorded in successive intervals of 30 minutes. A 30-minute interval was chosen as a fail-safe, and only resulted in the loss of a couple droplets in the transition from one interval to the next. Plots of $T_n$ against $n$ could be used to view the evolution of the drop interval as the system progressed through different regimes. Plots of the relationship between the successive drop time interval $T_n$ and the next interval $T_{n+1}$ were also created to study the evolution. Four selected regions demonstrating the four types of data encountered in this experiment are summarized in Fig. \ref{fig:tn_vs_tn1}.

In addition to the systematic drift and the missing droplet issue described in $\text{Fig. \ref{fig:tn_vs_tn1}}$, sharp jumps in the intervals between drops were seen throughout the data (visible most clearly in the rows two and three of Fig. \ref{fig:tn_vs_tn1}). These are not aligned with the starts of new 30-minute data collection intervals, and instead are currently attributed to local variations in the parameters affected the flow rate, which include changes in the water pressure, temperature, and density, and the presence of bubbles. Following these jumps the system repeats its previous pattern of dynamics, i.e. it is seen to reproduce the same evolution that it previously had at the same flow rate. 

\begin{figure}[!]
	\includegraphics[width=\columnwidth]{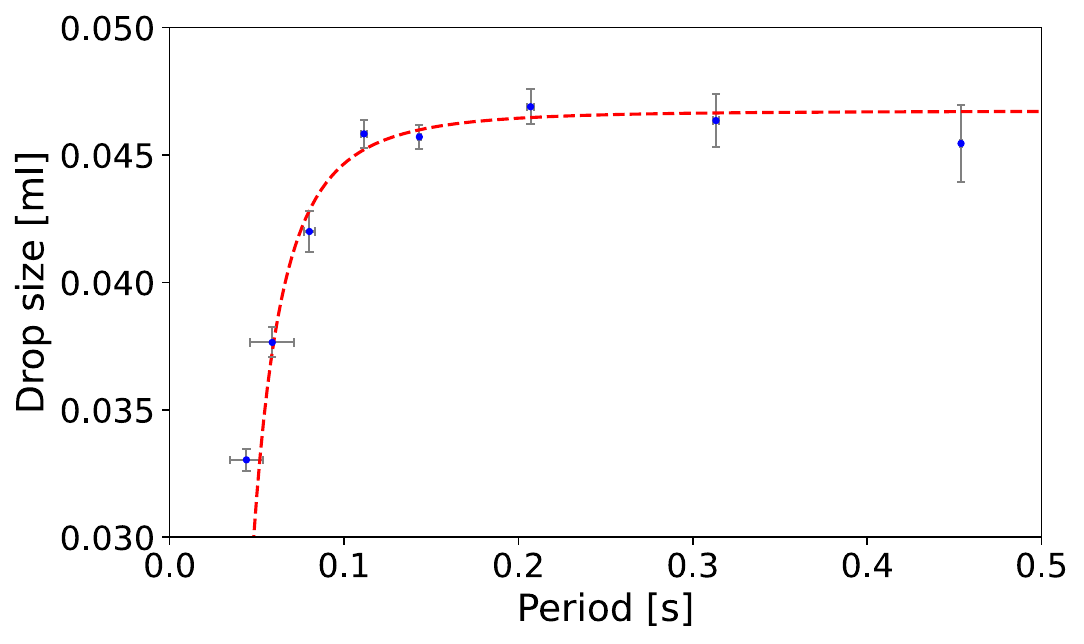}
    \caption{Water droplet size as a function of the interval of successive water droplets from a dripping faucet. A container filled with water was allowed to drain by dripping through a pipette opening. A valve connected to a stepper motor was used for the exploration of multiple drip rates. At each of the eight stepper motor settings plotted above, the time of each drop was recorded using an optical sensor for a short fixed interval $\Delta t$, the time taken to approximately fill a 10 ml graduated cylinder, which varied from 5 seconds to 30 seconds depending on the flow rate. A short time frame was chosen to minimize the effects of systematic drift from the decreasing water level in the tank. Recording the time of each drop provided a measure of the total number of drops, $N$, and the successive intervals between each $n$-th drop, $T_n$. The total volume of water collected $V_{\text{tot}}(N,\Delta t, T)$ was measured using a graduated cylinder. The drop interval $T$ at each point was calculated as the average of the intervals of each successive drop, with the uncertainty shown here as the standard deviation of the $N-1$ interval measurements. The drop size, $V_d(T)$, which varies with the drip interval $T$ as shown above, was calculated as $V_d(T) = V_{\text{tot}}/N$. The reading uncertainty in $V_d$ was estimated to be $\pm 0.1$ ml, based on reading the water level in the graduated cylinder based on the bottom of the meniscus. This was propagated to the uncertainty in drop size as $\delta V_d(T) = \pm 0.1 V_d / V_{\text{tot}}$ ml. The data was fit to $V_d(T) = AT^B +C$ as suggested in \cite{paper1} using orthogonal distance regression, where
    $A=(-28\pm7)\times10^{-7}\,\text{ml}$,
    $B=-2.8\pm0.9$, 
    $C=(467\pm5)\times10^{-4}\,\text{ml}$. The resulting reduced chi-square, $\chi_\nu^2$, of 0.83 indicates that the uncertainties may have been overestimated, or that the model is not a good fit for the data. The equation for drop size as a function of the successive drop interval, $V_d(T)$ (with units of ml per drop), will be used to scale a bifurcation diagram of the droplet flow rate (drops per min) into a volumetric flow rate (ml per min). This diagram will be produced by sampling the data shown in Fig. \ref{fig:tn_vs_tn1}.}
    \label{fig:drop_size}
\end{figure}

Before a bifurcation diagram could be created from the $10^6$ collected droplets, the relationship between the droplet size and period (averaged time between drops) had to be established. This was done by using a graduated cylinder to capture the amount of water collected in an interval for a particular (averaged) period. Short ($\sim$5--30 s) time intervals were used to minimize the effect of the aforementioned systematic period drift due to the tank emptying. A description of the calculation of the droplet size from data collected in this fashion is shown in Fig \ref{fig:drop_size}, which also shows the resulting plot of droplet size $V_d$ against averaged drop interval time $T$.

 Though averaging for one period $T$ across an interval meant that the variations of the period within that interval (for multi-period or chaotic regimes) were not accounted for, these were included in the form of a larger uncertainty in the reported ($T,V_d$) point.
 
\begin{figure}[!]
	\includegraphics[width=\columnwidth]{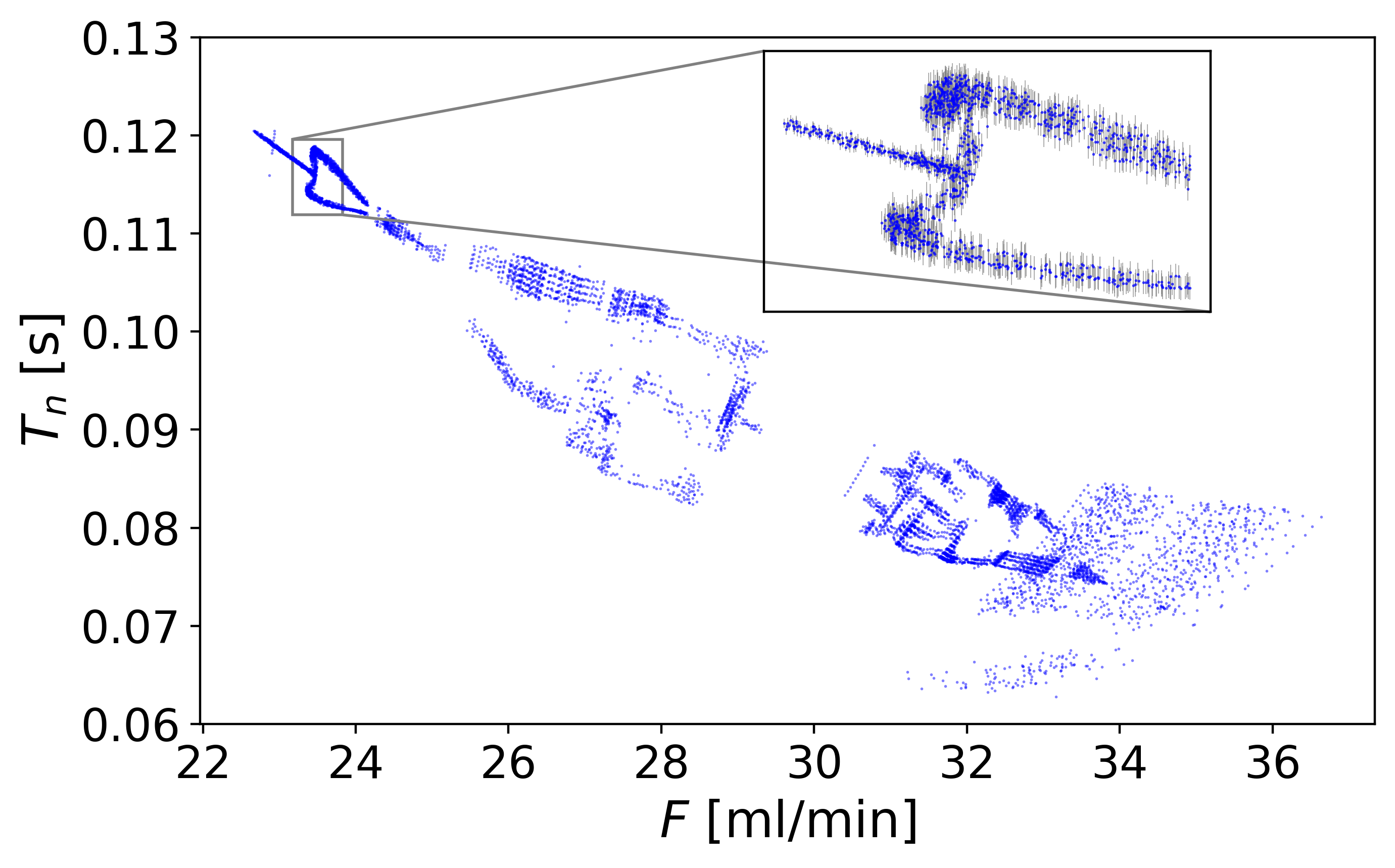}
    \caption{Bifurcation diagram showing the evolution of the successive time intervals $T_n$ of drops from a dripping water faucet through several dynamical regimes. A water tank was allowed to drain by dripping through an eyedropper, and the time of each drop was recorded. The decreasing water level in the tank decreased the pressure above the opening, and as a result the flow rate through the eyedropper, allowing the system to explore a continuous range of drip intervals from 0.02 s to 0.14 s. A subset of the data is shown here, with $T_n = [0.06,0.12]$ s. The $2.4\times10^5$ remaining drop time intervals were used to create points on this plot as follows: The data was discretized into sets of  $N=200$ points. Within each set, the systematic drift was accounted for by fitting the times between drops $T_n$ to a straight line, $T_n = \alpha n + \beta$, subtracting this fit from the points $T_n$, and adding the average time between drops in this interval, $T_{n,\text{avg}}$. A histogram distribution of the time intervals was made, with a bin size related to the elapsed time in the selected range, $\Delta t$, and the standard deviation of a single period regime, $\sigma_1$. The drop time intervals of bins with counts greater than the average by one sigma were selected as points to be plotted. The drop flow rate was calculated as $F_d=100/\Delta t$, and converted to a volumetric drop rate by via $F=F_dV_d(T)$, where $V_d(T)$ is obtained from the fit in Fig. \ref{fig:drop_size} and $T$ is the average period in that 200-point interval.
    The inset shows a section of the plot with a transition from period-1 to period-2 dripping with vertical uncertainty bars, taken as the width of a bin on the $T_n$ histogram. The uncertainty in flow rate is not shown, but is estimated to be substantial, on the order of 3 ml/min, based on propagating the standard deviation of the bin count through the flow rate calculation, with uncertainties in the drop to volume conversion included. Behavior contradicting the expected successive transitions from period-$n$ behavior with $n=1,2,4,...$ eventually leading to chaos is observed. A return from period-2 to period-1 is observed, along with complex internal self-similarities, before an eventual transition to chaos past a flow rate of 34 ml/min. The collection of more data sets would show whether this is the result of external environmental contributions or a property of the dripping water faucet chaotic system.}
    \label{fig:bifurcation}
\end{figure}

 All of the ingredients are now available for creating a bifurcation diagram. A subset of the dataset partially shown in Fig. \ref{fig:tn_vs_tn1} was used in order to remove the missing droplet effect. The bifurcation diagram and a discussion of its calculation is shown in Fig. \ref{fig:bifurcation}.

To compliment the figure's brief summary, the uncertainty propagation was performed as follows: In the vertical direction, simply the width of a bin was taken. Although a full bin width is taken, rather than the standard deviation of the points making up that bin, this is still expected to be an underestimate, as it does not account for the linear systematic drift correction, nor the sudden jumps in the drop interval data discussed in \ref{fig:tn_vs_tn1}. The uncertainty in flow rate was propagated as
\begin{equation}
    \delta F = F \sqrt{\lr{\frac{\sigma_b}{100}}^2 + \lr{\frac{\delta_T}{\Delta t}}^2} \delta V_d 
    \approx
    F \frac{\sigma_b}{100}  \delta V_d(T) \sim 3\,\frac{\text{ml}}{\text{min}},
\end{equation}
where $\sigma_b$ is the standard deviation of the counts in the histogram distribution and $\delta_T$ is the width of a bin in the $T_n$ distribution, and $\delta V_d(T)$ is the uncertainty associated with the conversion from a drop to a volumetric flow rate, following the equation in Fig. \ref{fig:drop_size}. Though this represents an error of about 10$\%$ on the flow rate, there are apparent details and complex patterns in Fig. \ref{fig:bifurcation} which have a resolution on the order of $10^{-1}$ ml/min. This indicates that the uncertainty propagation for the uncertainty in flow rate is overestimated.

One general point to note is that Fig. \ref{fig:bifurcation} shows an overall downward trend, since the interval between successive drops $T_n$ decreases with increasing flow rate, whereas the left column of Fig. \ref{fig:tn_vs_tn1} shows an upward trend, since the interval $T_n$ is increasing as the tank empties itself and the flow rate decreases.

As mentioned in Fig. \ref{fig:bifurcation}, the system appears to be exhibiting complex self-similarities, or patterns within patterns. This is seen in the flow rate region $[25,29]$ ml/min in Fig. \ref{fig:bifurcation}, but perhaps more clearly illustrated in the second row of Fig. \ref{fig:tn_vs_tn1}. The behavior is not entirely bifurcated into four periods as it is into two in row 3. Instead, there is a region, characterized by an upper and lower limit on $T_n$, within which a scatter of points is distributed. Without looking closely, one could interpret this as purely chaotic behavior. However, the overlaying of semi-transparent points as is done in the figure shows that there are at most clearly four branches within this region. Within these branches, a bifurcation can also be observed. 

These results show a contradiction with the numerical predictions in \cite{dinnocenzo_1996_dripping}: the system is not observed to linearly transition between distinct increasingly chaotic regimes with increasing flow rate. In other words, it is not seen to change from single period dripping, to period-$m$ dripping, to pure chaos in a successive fashion.
There is more complexity to the system that expected from numerical simulations. The resulting bifurcation diagram therefore does not allow the Feigenbaum constants to be estimated, since it does not show the typical successive transitions seen on a standard logistical map. A qualitatively ``clean'' transition from a period-2 regime to a period-4 regime, required for (1), cannot be determined from Fig. \ref{fig:bifurcation}. 

As a test of reproducibility, the analysis was repeated on a smaller dataset with a more constrained range of dynamics, and a similar pattern was obtained---complex self-similarities and multiple transitions between period-2 and period-4 regimes were observed. 

In short, this letter studies the behavior of the period in a seemingly simple system, a dripping faucet. The system was initially expected to traverse the standard dynamical regimes associated with chaos---a transition from simple (period-1) dripping to complex dripping with period-$n$, and finally to chaotic responses. Namely, it was expected to do so in successive fashion with increasing flow rate, without returning to previously visited regimes. This would be in accordance with numerical simulations \cite{chaos_manual_paper}. However, a bifurcation diagram created from observing the evolution of the system resulting from a decreasing flow rate, and calibrated to account for varying droplet size, showed that the system exhibits much more intricate behavior. Specifically, for an increasing flow rate, the dripping faucet revisited regimes with fewer branches, and was observed with complex inner bifurcations and semi-chaotic regimes. This prevented the calculation of Feigenbaum constants, which require a well-defined bifurcation diagram with successive transitions through chaotic regimes. Further data collection and analysis would be required to provide insight into the nature of these complex patterns, by checking their reproducibility. A more detailed study of the chaotic dynamics of a dripping water faucet could be the gateway to understanding other even more complex chaotic systems.

\pagebreak










\bibliography{mybiblio}

\end{document}